\documentclass[12pt,]{article}
 \usepackage{graphicx}
 \usepackage[cp1251]{inputenc}  
 \usepackage[russian]{babel}    %
\begin{document}

\begin{center}{\large\bf VARIATIONS OF ANGULAR MOMENTUM $L_z$ AS AN INDICATOR OF ORBITAL CHAOS OF GLOBULAR CLUSTERS IN THE CENTRAL REGION  OF THE GALAXY WITH A BAR}
\end{center}
  \bigskip
\begin{center}
A. T.~Bajkova \footnote{e-mail:bajkova@gaoran.ru}, A. A.~Smirnov, V. V.~Bobylev
\end{center}
\begin{center}
{\small\it  Central Astronomical Observatory of the Russian Academy of Sciences, Pulkovo }
\end{center}

 \bigskip

It is shown how the violation of the invariance of the $Z$-component of the orbital angular momentum $L_z$ in the axially symmetric potential of the Galaxy with a bar can serve as an indicator of the degree of orbital chaos of globular clusters in the central region of the Galaxy. In this case, the higher the variations of $L_z$ of the orbit over a certain period of time, the higher the chaos of the orbit. In essence, a new method for analyzing orbital dynamics -- regular or chaotic -- is proposed. A high level of correlation between the results of orbit classification by the proposed method and the results of classification by other methods is shown. As a result, a sample of 45 globular clusters in the central region of the Galaxy with a radius of 3.5 kpc is divided into regular, chaotic, and weakly chaotic.

\bigskip\noindent
\noindent {\it Keywords:} Galaxy, bar, globular clusters, chaotic orbital dynamics

\newpage

\section*{INTRODUCTION}

This work is essentially a continuation of the works~ [1-3], devoted to the study of the orbital dynamics --  regular or chaotic -- of globular clusters in the central region of the Galaxy. As in previous works, the sample includes 45 globular clusters in the central region of the Galaxy with a radius of 3.5 kpc. To form the 6D phase space required for integrating the orbits, the most accurate astrometric data to date from the Gaia satellite~ [4], as well as new refined average distances~ [5] were used. The following, the most realistic parameters of the bar, known from the literature~ [6,7] were adopted: mass $10^{10} M_\odot$, length of the major semi-axis 5 kpc, angle of rotation of the bar axis 25$^o$, rotation velocity 40 km s$^{-1}$ kpc$^{-1}$.

The aim of this work is to establish a connection between the variations of the $Z$-component of the angular momentum of the orbit $L_z$ in time with the degree of chaotization of the orbit of the GCs of our sample in the potential with the adopted central rotating bar, in which the value $L_z$ is no longer an invariant, as in the axisymmetric potential.

The paper is structured as follows. The first section provides a brief description of the adopted potential models—an axisymmetric potential and a nonaxisymmetric potential including a bar. The second section provides references to the astrometric data used, as well as the method for forming the GC sample. The third section describes the proposed method for estimating the regularity/chaoticity of motion based on calculating the variation of $L_z$ over a given period of time. The fourth section analyzes the obtained results and establishes the relationship between the new approach and the methods proposed and investigated in earlier studies. The CONCLUSIONS formulate the main results of the paper.

\section{GALACTIC POTENTIAL MODEL}

\subsection{Axisymmetric potential}

The axisymmetric gravitational potential of the Galaxy, traditionally used by us (see, for example,~[2]) for integrating the GC orbits, is represented as the sum of three components~--- the central spherical
bulge $\Phi_b(r(R,Z))$, the disk $\Phi_d(r(R,Z))$ and the massive
spherical halo of dark matter $\Phi_h(r(R,Z))$:
 \begin{equation}
 \begin{array}{lll}
  \Phi(R,Z)=\Phi_b(r(R,Z))+\Phi_d(r(R,Z))+\Phi_h(r(R,Z)).
 \label{pot}
 \end{array}
 \end{equation}
Here we use a cylindrical coordinate system ($R,\psi,Z$) with the origin at the galactic center. In a rectangular coordinate system $(X,Y,Z)$ with the origin at the galactic center, the distance to a star (spherical radius) will be $r^2=X^2+Y^2+Z^2=R^2+Z^2$, where the $X$ axis is directed from the Sun to the galactic center, the $Y$ axis is perpendicular to the $X$ axis in the direction of galactic rotation, and the $Z$ axis is perpendicular to the galactic plane $(X,Y)$ toward the north galactic pole. The gravitational potential is expressed in units of 100 km$^2$/s$^2$, distances~--- in kpc, masses~--- in units of the galactic mass $M_{gal}=2.325\times 10^7 M_\odot$, corresponding to the gravitational constant $G=1$.

The axisymmetric potentials of the bulge $\Phi_b(r(R,Z))$ and disk $\Phi_d(r(R,Z))$
are represented in the form proposed in the work~[8]:
 \begin{equation}
  \Phi_b(r)=-\frac{M_b}{(r^2+b_b^2)^{1/2}},
  \label{bulge}
 \end{equation}
 \begin{equation}
 \Phi_d(R,Z)=-\frac{M_d}{\Biggl[R^2+\Bigl(a_d+\sqrt{Z^2+b_d^2}\Bigr)^2\Biggr]^{1/2}},
 \label{disk}
\end{equation}
where $M_b, M_d$~ are the masses of the components, $b_b, a_d, b_d$~ are the scale parameters of the components in kiloparsecs. The halo component (NFW) is represented according to the work~[9]:
 \begin{equation}
  \Phi_h(r)=-\frac{M_h}{r} \ln {\Biggl(1+\frac{r}{a_h}\Biggr)},
 \label{halo-III}
 \end{equation}
where $M_h$ is the mass, $a_h$ is the scale parameter.
Table 1 shows the parameter values ??of the adopted model
of the galactic potential.

\begin{figure*}
{\begin{center}

\includegraphics[width=0.4\textwidth,angle=-90]{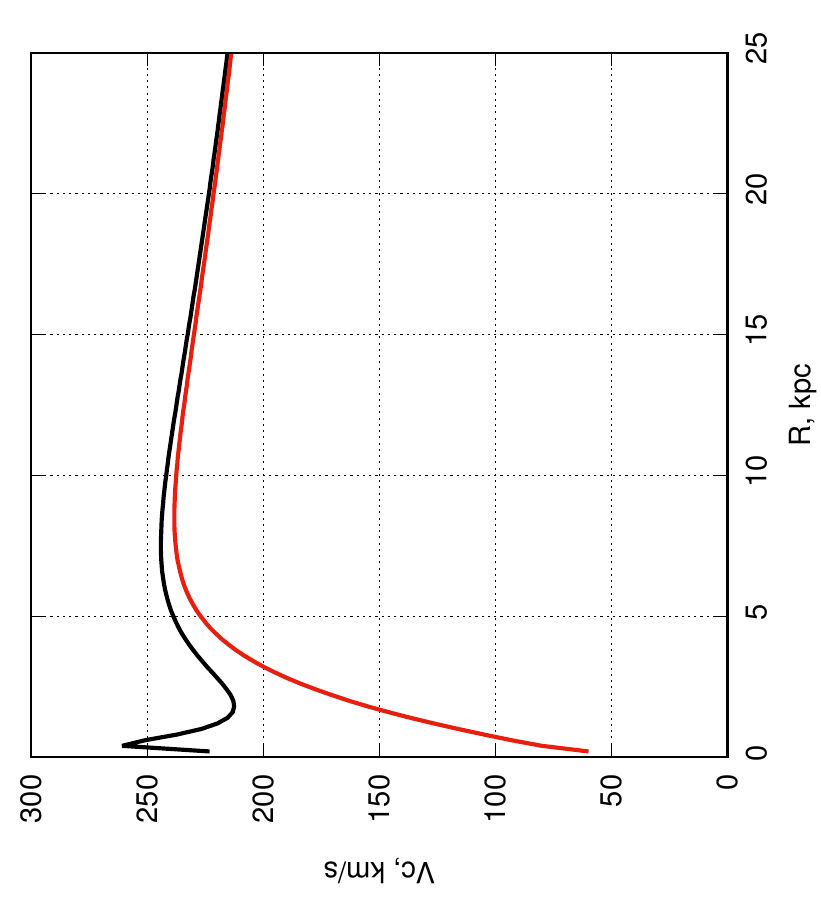}

\bigskip

\caption{\small Rotation curve of the Galaxy with an axisymmetric potential without a bar (black line) and a non-axisymmetric potential including a bar (red line).}
\label{fcomp}
\end{center}}
\end{figure*}

\subsection{Модель бара}

The model of a three-axis ellipsoid was chosen as the central bar potential~[6]:
\begin{equation}
  \Phi_{bar} = -\frac{M_{bar}}{(q_b^2+X^2+[Ya/b]^2+[Za/c]^2)^{1/2}},
\label{bar}
\end{equation}
where $X=R\cos\vartheta, Y=R\sin\vartheta$, $a, b, c$~--- are the three
semi-axes of the bar, $q_b$~--- is the bar's scale parameter (the length of the bar's major semi-axis);
$\vartheta=\theta-\Omega_{b}t-\theta_{b}$, $tg(\theta)=Y/X$,
$\Omega_{b}$~--- is the circular velocity of the bar, $t$~--- is the integration time, $\theta_{b}$~--- is the orientation angle of the bar
relative to the galactic axes $X,Y$, measured from the line
connecting the Sun and the galactic center (the $X$ axis) to the major axis of the bar in the direction of galactic rotation.

Based on information in numerous literature, in particular in [6], the following bar parameters were used: $M_{bar}=430 M_{gal}$, $\Omega_{b}=$40 km s$^{-1}$ kpc$^{-1}$, $q_b=5$ kpc, $\theta_{b}=25^o$. The adopted bar parameters are listed in Table~1.

 {\begin{table}[t]                                    
 \caption[]
 {\small\baselineskip=1.0ex
Values ??of the parameters of the galactic potential model, $M_{gal}=2.325\times 10^7 M_\odot$
  }
 \label{t:model-III}
 \begin{center}\begin{tabular}{|c|r|}\hline
 $M_b$ &   443 M$_{gal}$ \\
 $M_d$ &  2798 M$_{gal}$ \\
 $M_h$ & 12474 M$_{gal}$ \\
 $b_b$ & 0.2672 kpc  \\
 $a_d$ &   4.40 kpc  \\
 $b_d$ & 0.3084 kpc  \\
 $a_h$ &    7.7 kpc  \\
\hline\hline
 $M_{bar}$ & 430 M$_{gal}$ \\
 $\Omega_b$ & 40 km s$^{-1}$ kpc$^{-1}$ \\
 $q_b$     &  5.0 kpc  \\
 $\theta_{b}$ &  $25^o$   \\\hline
 $a/b$ & 2.38  \\
 $a/c$ & 3.03  \\
    \hline
 \end{tabular}\end{center}\end{table}}

To integrate the equations of motion, we used the fourth-order Runge-Kutta algorithm.

The value of the peculiar velocity of the Sun relative to the local standard of rest was taken to be equal to
$(u_\odot,v_\odot,w_\odot)=(11.1,12.2,7.3)\pm(0.7,0.5,0.4)$~km s$^{-1}$
according to the work~[10]. The elevation of the Sun above the plane of the Galaxy is taken to be 16 pc in accordance with the work [11].

For comparison, Fig. 1 shows the obtained model rotation curves: an axisymmetric potential (black line) and a potential with a bar (red line).

\section{DATA}

Data on the proper motions of GCs are taken from a new catalog [4], compiled based on Gaia EDR3 observations. The average distances to globular clusters are taken from [5].

Our catalog of GCs [12] contains 152 objects.
The selection of globular clusters from this set, belonging to the bulge/bar region, was carried out in accordance with a purely geometric criterion, discussed in [13], and also used by us in [14]. This criterion is very simple and consists of selecting GCs whose apocentric distance from their orbits does not exceed the bulge radius, which is usually taken to be 3.5 kpc. The orbits are calculated in an axisymmetric potential. The full list of 45 objects in our sample is presented in Table 2, which presents the results of the analysis of the chaoticity/regularity of the GC orbits (the first column gives the GC serial number, the second the GC name).
\label{lab:text1}

\section{METHOD}

The method consists of calculating the variation (from the minimum to the maximum value) of the angular momentum change  {$L_z$} in a potential with a bar over a certain, fairly long time interval (for clarity), divided by the modulus of  {$L_z$} at the initial moment in time for the orbit of each GC. The desired variation for GC number $n=1,...,45$  is denoted as
 {$D(L^n_z/|L^n_z(1)|)$} . In our case, as in other previously reviewed works~[1--3], the integration time of the GC orbit is 120 billion years. A threshold value is then established for deciding whether chaos is present. For this purpose, the entire set of variations (45 GCs) is analyzed, histograms, etc., are constructed, and correlations with other known methods of orbital dynamics analysis are established.

\section{RESULTS}

The results of applying the above method to the entire sample of 45 GCs are presented in column 10 of Table 2. Column 11 provides the verdict: chaotic (C) or regular (R) orbit based on the voting method, taking into account both new results and the results of applying other methods for analyzing GC dynamics [1,3], reflected in columns 3 through 9. All method names listed in the table header are numbered from 1 to 9. That is, each method has its own number. In addition, the bibliographic source is given in square brackets. Table 3 presents the correlation coefficients between the classification results of GC dynamics (regular or chaotic) obtained by the presented methods in pairs (i.e., it essentially represents a correlation matrix).
The correlation coefficients allow one to judge the similarity of the classification results obtained by different methods. We see that a significant correlation has been obtained between the new method and other methods.

A graphical representation of the obtained results is given in Fig. 2. Figure (a) shows the variations of $D(L^n_z/|L^n_z(1)|)$ (parallel to the abscissa) for each GC in accordance with the Jacobi integral, which is an invariant for the orbit of each GC (ordinate axis). The variations for GCs with chaotic dynamics determined by the new method are shown in blue, and for GCs with regular dynamics, in red. We see that the blue lines are much longer compared to the red ones. We determined the variation threshold value that divides the orbits into chaotic and regular from the histogram shown in Figure (b). This number represents $\sqrt(2)$.
The numerical values ??of the variations of $D(L^n_z/|L^n_z(1)|)$ for all 45 GCs in our sample are shown in Figure (c). The histogram (b) is obtained precisely from the distribution (c). According to these graphs, the regular part of the histogram is displayed by the first peak, equal to 25 ns, for which the variations are $D(L^n_z/|L^n_z(1)|)\leq \sqrt(2)$.

Figure (d) shows the degree of chaos or regularity obtained by "voting"
all the methods listed in Table 2. The closer the score to 1, the more regular the orbit. The closer the score to 0, the more chaotic the orbit. Orbits with a score of 1 belong to the following GCs with regular dynamics: NGC6266, Terzan4, Liller1, Terzan1, Terzan6, Terzan9, NGC6522, NGC6624, NGC6637, NGC6717, Pismi26, NGC6569, E456-78, NGC6540, Djorg2, NGC6171, NGC6539, NGC6553. Orbits with a score of ``0'' belong to the following GCs with chaotic dynamics: E452-11, NGC6273, NGC6293, NGC6342, NGC6355, Terzan2, BH229, NGC6401, Pal6, NGC6453, NGC6558, NGC6626, NGC6638, NGC6642, NGC6652. The first group consists of 18 GCs, the second -- 15 GCs. The remaining GCs with intermediate values ??of the degree of regularity make up a group of 12 GCs -- NGC6144, NGC6380, Terzan5, NGC6440, NGC6528, NGC6723, Terzan3, NGC6256, NGC6304, NGC6325, NGC6316, NGC6388, which we designated as weakly chaotic (WC).

\section*{CONCLUSIONS}

\medskip

The following main results were obtained:

1. It is shown how the violation of the invariance of the $Z$-component of the orbital angular momentum $L_z$ in the nonaxisymmetric potential of the Galaxy with a bar can serve as an indicator of the degree of chaos of globular clusters in the central region of the Galaxy. Moreover, the higher the variations of the orbit $L_z$ over a given period of time, the greater the orbital chaos. Thus, a new method for analyzing the orbital dynamics of GCs and determining their degree of chaos is proposed. A significant correlation between the proposed method and other methods is obtained, demonstrating its good performance.

2. By comparing the proposed method with other methods of analysis using the "voting" method, a sample of 45 globular clusters within a radius of 3.5 kpc was divided into regular, chaotic, and weakly chaotic.
The following lists of GCs were obtained:

2.1. List of 18 globular clusters with regular dynamics (R):

NGC6266, Terzan4, Liller1, Terzan1, Terzan6, Terzan9, NGC6522, NGC6624, NGC6637, NGC6717, Pismi26, NGC6569, E456-78, NGC6540, Djorg2, NGC6171, NGC6539, NGC6553.

2.2. List of 15 globular clusters with chaotic dynamics (C):
E452-11, NGC6273, NGC6293, NGC6342, NGC6355, Terzan2, BH229, NGC6401, Pal6, NGC6453, NGC6558, NGC6626, NGC6638, NGC6642, NGC6652.

2.3. List of 12 globular clusters with weak chaotic dynamics (WC):

NGC6144, NGC6380, Terzan5, NGC6440, NGC6528, NGC6723, Terzan3, NGC6256, NGC6304, NGC6325, NGC6316, NGC6388

\newpage
\noindent{\bf References}

\bigskip

\begin{enumerate}

 \item
A. T. Bajkova, A. A. Smirnov, V. V. Bobylev. Publications of the Pulkovo Observatory  {\bf 233}, 1 (2024) DOI:10.31725/0367-7966-2024-233-1-28, arXiv: 2406.15590.

 \item
A. T. Bajkova, A. A. Smirnov, V. V. Bobylev. Publications of the Pulkovo Observatory {\bf 235}, 1 (2024) DOI:10.31725/0367-7966-2024-235-1-15, arXiv: 2412.02426.

 \item
A. T. Bajkova, A. A. Smirnov, V. V. Bobylev. Publications of the Pulkovo Observatory {\bf 236}, 1 (2025) DOI:10.31725/0367-7966-2025-236-1-22.


\item
E. Vasiliev, H. Baumgardt. Monthly Notices of the Royal Astronomical Society {\bf 505}, Issue 4, 5978 (2021), arXiv: 2102.09568.

\item
H. Baumgardt, E. Vasiliev. Monthly Notices of the Royal Astronomical Society {\bf 505}, Issue 4, 5957 (2021), arXiv: 2105.09526.

\item
J. Palou$\breve{s}$, B. Jungwiert, J. Kopeck$\acute{y}$. Astronomy and Astrophysics {\bf 274}, 189 (1993).

\item
J. L. Sanders, L. Smith, N. W. Evans, P. Lucas. Monthly Notices of the Royal Astronomical Society  {\bf 487}, Issue 4, 5188 (2019), arXiv: 1903.02008.

\item
M. Miyamoto, R. Nagai, PASJ {\bf 27}, 533 (1975).

\item
J.F. Navarro, C.S. Frenk, S.D.M. White, Astrophys. J. {\bf 490}, 493 (1997).

\item
R. Sch\"onrich, J. Binney, W. Dehnen, Monthly Notices of the Royal Astronomical Society {\bf 403}, 1829 (2010).

\item
V.V. Bobylev, A.T. Bajkova, Astron. Lett. {\bf 42}, 1 (2016)

\item
A. T. Bajkova, V. V. Bobylev. Publications of the Pulkovo Observatory {\bf 227}, 1 (2022)
DOI: 10.31725/0367-7966-2022-227-2, arXiv: 2212.00739.

\item
D. Massari, H.H. Koppelman, A. Helmi, Astron. Astrophys. {\bf 630}, L4 (2019).

\item
A. T. Bajkova, G. Carraro, V.I. Korchagin, N.O. Budanova, V.V. Bobylev,
Astrophys. J. {\bf 895}, 69 (2020).

\end{enumerate}

{\begin{table*}[t]                                    
 \caption[]
 {\baselineskip=1.0ex
Summary table of signs of regularity (R) and chaos (C) of the orbits of 45 GCs.
  }
 \label{t:f}
 {\scriptsize\begin{center}\begin{tabular}{|r|l||c|c|c|c|c|c|c|c|c|}\hline
    &        & Probabi-&MEGNO    & Maximum      &Poincare&Frequency  &Visual&New method:& New method & Voting \\
 №  &Name& listic&    & Lyapunov    &sections  &drift &assess- & entropy &  &method  \\
    &of GC & method  & &exponents &  &$\lg(\Delta f)$&ment &estimate&  $L_z/L_z(1)$ &\\
    &        &   (1)  [1] & (2)  [1]   & (3)  [1]    & (4)  [1] & (5)  [1]          & (6)  [1]  & (7) [3]& (8) this work & (9)       \\\hline
 1  &NGC6144 & (C)    & 2.173  (R)& -0.002  (R)& (R)   &-2.08 (R)&  (R)  &  0.021 (R)& 1.33 (R)&7/8 (WC) \\\hline
 2  &E452-11 & (C)    & 0.752  (C)& 0.919   (C)& (C)   &-1.37 (C)&  (C)  &  0.215 (C)& 11.02(C)&0 (C) \\\hline
 3  &NGC6266 & (R)    & 1.976  (R)& -0.017  (R)& (R)   &-4.00 (R)&  (R)  &  0.009 (R)& 1.19 (R)&1 (R) \\\hline
 4  &NGC6273 & (C)    & 1.494  (C)& 1.318   (C)& (C)   &-1.77 (C)&  (C)  &  0.050 (C)& 4.39 (C)&0 (C) \\\hline
 5  &NGC6293 & (C)    & 0.934  (C)& 4.167   (C)& (C)   &-0.07 (C)&  (C)  &  0.063 (C)& 2.81 (C)&0 (C) \\\hline
 6  &NGC6342 & (C)    & 0.769  (C)& 0.428   (C)& (C)   &-2.14 (C)&  (C)  &  0.095 (C)& 1.75 (C)&0 (C) \\\hline
 7  &NGC6355 & (C)    & 0.509  (C)& 2.257   (C)& (C)   &-0.10 (C)&  (C)  &  0.123 (C)& 6.87 (C)&0 (C) \\\hline
 8  &Terzan2 & (C)    & 0.627  (C)& 0.905   (C)& (C)   &-0.23 (C)&  (C)  &  0.224 (C)& 2.09 (C)&0 (C) \\\hline
 9  &Terzan4 & (R)    & 1.993  (R)& -0.144  (R)& (R)   &-4.00 (R)&  (R)  &  0.010 (R)& 1.17 (R)&1 (R) \\\hline
10  &BH229   & (C)    & 0.663  C& 2.220     (C)& (C)   &-1.81 (C)&  (C)  &  0.108 (C)& 10.22(C)&0 (C) \\\hline
11  &Liller1 & (R)    & 2.049  (R)& -0.037  (R)& (R)   &-4.00 (R)&  (R)  &  0.010 (R)& 1.18 (R)&1 (R) \\\hline
12  &NGC6380 & (R)    & 2.182  (R)& 0.220   (C)& (R)   &-3.72 (R)&  (R)  &  0.010 (R)& 1.27 (R)&7/8 (WC) \\\hline
13  &Terzan1 & (R)    & 2.000  (R)& -0.029  (R)& (R)   &-4.00 (R)&  (R)  &  0.008 (R)& 1.09 (R)&1 (R) \\\hline
14  &NGC6401 & (C)    & 0.622  (C)& 4.712   (C)& (C)   &-0.09 (C)&  (C)  &  0.134 (C)& 8.11 (C)&0 (C) \\\hline
15  &Pal6    & (C)    & 0.502  (C)& 3.359   (C)& (C)   &-0.10 (C)&  (C)  &  0.131 (C)& 11.57(C)&0 (C) \\\hline
16  &Terzan5 & (R)    & 2.023  (R)& 0.041   (C)& (R)   &-4.00 (R)&  (R)  &  0.009 (R)& 1.33 (R)&7/8 (WC) \\\hline
17  &NGC6440 & (C)    & 1.901  (R)& 0.572   (C)& (R)   &-2.26 (C)&  (C)  &  0.038 (C)& 1.68 (C)&2/8 (WC) \\\hline
18  &Terzan6 & (R)    & 1.996  (R)& -0.055  (R)& (R)   &-4.00 (R)&  (R)  &  0.007 (R)& 1.06 (R)&1 (R) \\\hline
19  &NGC6453 & (C)    & 1.178  (C)& 1.998   (C)& (C)   &-1.92 (C)&  (C)  &  0.170 (C)& 4.13 (C)&0 (C) \\\hline
20  &Terzan9 & (R)    & 2.358  (R)& -0.056  (R)& (R)   &-3.86 (R)&  (R)  &  0.012 (R)& 1.23 (R)&1 (R) \\\hline
21  &NGC6522 & (R)    & 1.996  (R)& -0.020  (R)& (R)   &-4.00 (R)&  (R)  &  0.013 (R)& 1.09 (R)1 &(R) \\\hline
22  &NGC6528 & (R)    & 2.008  (R)& -0.036  (R)& (R)   &-4.00 (R)&  (R)  &  0.015 (R)& 2.00 (C)&7/8 (WC) \\\hline
23  &NGC6558 & (C)    & 0.819  (C)& 1.364   (C)& (C)   &-1.03 (C)&  (C)  &  0.099 (C)& 6.53 (C)&0 (C) \\\hline
24  &NGC6624 & (R)    & 1.847  (R)& -0.040  (R)& (R)   &-4.00 (R)&  (R)  &  0.008 (R)& 1.09 (R)&1 (R) \\\hline
25  &NGC6626 & (C)    & 1.194  (C)& 0.093   (C)& (C)   &-1.78 (C)&  (C)  &  0.079 (C)& 2.31 (C)&0 (C) \\\hline
26  &NGC6638 & (C)    & 0.533  (C)& 2.411   (C)& (C)   &-0.16 (C)&  (C)  &  0.140 (C)& 7.66 (C)&0 (C) \\\hline
27  &NGC6637 & (R)    & 1.988  (R)& -0.012  (R)& (R)   &-4.00 (R)&  (R)  &  0.008 (R)& 1.28 (R)&1 (R) \\\hline
28  &NGC6642 & (C)    & 0.681  (C)& 2.451   (C)& (C)   &-1.01 (C)&  (C)  &  0.197 (C)& 2.24 (C)&0 (C) \\\hline
29  &NGC6717 & (R)    & 2.044  (R)& -0.001  (R)& (R)   &-4.00 (R)&  (R)  &  0.015 (R)& 1.08 (R)&1 (R) \\\hline
30  &NGC6723 & (R)    & 2.252  (R)& 0.064   (C)& (R)   &-4.00 (R)&  (R)  &  0.013 (R)& 2.01 (C)&6/8 (WC) \\\hline
31  &Terzan3 & (R)    & 3.495  (R)& -0.000  (R)& (R)   &-1.89 (R)&  (R)  &  0.031 (C)& 1.44 (C)&6/8 (WC) \\\hline
32  &NGC6256 & (C)    & 0.893  (C)& -0.000  (R)& (C)   &-1.93 (C)&  (C)  &  0.039 (C)& 1.37 (R)&2/8 (WC) \\\hline
33  &NGC6304 & (R)    & 1.753  (R)& -0.000  (R)& (C)   &-1.38 (C)&  (C)  &  0.017 (R)& 1.27 (R)&5/8 (WC) \\\hline
34  &Pismi26 & (R)    & 1.941  (R)& -0.000  (R)& (R)   &-4.00 (R)&  (R)  &  0.009 (R)& 1.34 (R)&1 (R) \\\hline
35  &NGC6569 & (R)    & 1.957  (R)& -0.000  (R)& (R)   &-4.00 (R)&  (R)  &  0.009 (R)& 1.05 (R)&1 (R) \\\hline
36  &E456-78 & (R)    & 1.983  (R)& -0.000  (R)& (R)   &-3.59 (R)&  (R)  &  0.010 (R)& 1.19 (R)&1 (R) \\\hline
37  &NGC6540 & (R)    & 1.999  (R)& -0.000  (R)& (R)   &-4.00 (R)&  (R)  &  0.010 (R)& 1.05 (R)&1 (R) \\\hline
38  &NGC6325 & (C)    & 1.216  (C)& -0.000  (R)& (C)   &-3.22 (R)&  (C)  &  0.037 (C)& 1.47 (C)&2/8 (WC) \\\hline
39  &Djorg2  & (R)    & 2.320  (R)& -0.050  (R)& (R)   &-4.00 (R)&  (R)  &  0.009 (R)& 1.06 (R)&1 (R) \\\hline
40  &NGC6171 & (R)    & 2.015  (R)& -0.000  (R)& (R)   &-4.00 (R)&  (R)  &  0.009 (R)& 1.13 (R)&1 (R) \\\hline
41  &NGC6316 & (R)    & 2.289  (R)& 0.251   (C)& (R)   &-1.96 (R)&  (R)  &  0.035 (C)& 1.41 (R)&6/8 (WC) \\\hline
42  &NGC6388 & (R)    & 2.450  (R)& 0.271   (C)& (C)   &-0.03 (C)&  (R)  &  0.034 (C)& 1.07 (R)&4/8 (WC) \\\hline
43  &NGC6539 & (R)    & 1.993  (R)& -0.000  (R)& (R)   &-4.00 (R)&  (R)  &  0.009 (R)& 1.14 (R)&1 (R) \\\hline
44  &NGC6553 & (R)    & 1.899  (R)& -0.000  (R)& (R)   &-4.00 (R)&  (R)  &  0.008 (R)& 1.07 (R)&1 (R) \\\hline
45  &NGC6652 & (C)    & 1.121  (C)& 3.269   (C)& (C)   &-0.12 (C)&  (C)  &  0.149 (C)& 14.02(C)&0 (C) \\\hline
 \end{tabular}\end{center}}\end{table*}}

 {\begin{table*}[t]                                    
 \caption[]
 {\baselineskip=1.0ex
Correlation matrix
  }
 \label{t:f2}
 {\begin{center}\begin{tabular}{|r|c|c|c|c|c|c|c|c|c|}\hline
Метод&  (1)     & (2)  & (3)  & (4)     & (5)       & (6)       & (7)  &  (8)  &  (9) \\\hline
(1)  &  1.00    & 0.91 & 0.64 & 0.82    & 0.82      &  0.91     & 0.82 & 0.77  & 0.91 \\\hline
(2)  &  0.91    & 1.00 & 0.65 & 0.91    & 0.82      &  0.91     & 0.83 & 0.78  & 0.91 \\\hline
(3)  &  0.64    & 0.65 & 1.00 & 0.64    & 0.73      &  0.64     & 0.73 & 0.69  & 0.73 \\\hline
(4)  &  0.82    & 0.91 & 0.64 & 1.00    & 0.91      &  0.91     & 0.82 & 0.68  & 0.91 \\\hline
(5)  &  0.82    & 0.82 & 0.73 & 0.91    & 1.00      &  0.91     & 0.82 & 0.68  & 0.91 \\\hline
(6)  &  0.91    & 0.91 & 0.64 & 0.91    & 0.91      &  1.00     & 0.82 & 0.77  & 0.91 \\\hline
(7)  &  0.82    & 0.83 & 0.73 & 0.82    & 0.82      &  0.82     & 1.00 & 0.78  & 0.91 \\\hline
(8)  &  0.77    & 0.78 & 0.69 & 0.68    & 0.68      &  0.77     & 0.78 & 1.00  & 0.77 \\\hline
(9)  &  0.91    & 0.91 & 0.73 & 0.91    & 0.91      &  0.91     & 0.91 & 0.77  & 1.00 \\\hline
\end{tabular}\end{center}}\end{table*}}
\newpage

\begin{figure*}
{\begin{center}

        \includegraphics[width=1.0\textwidth,angle=0]{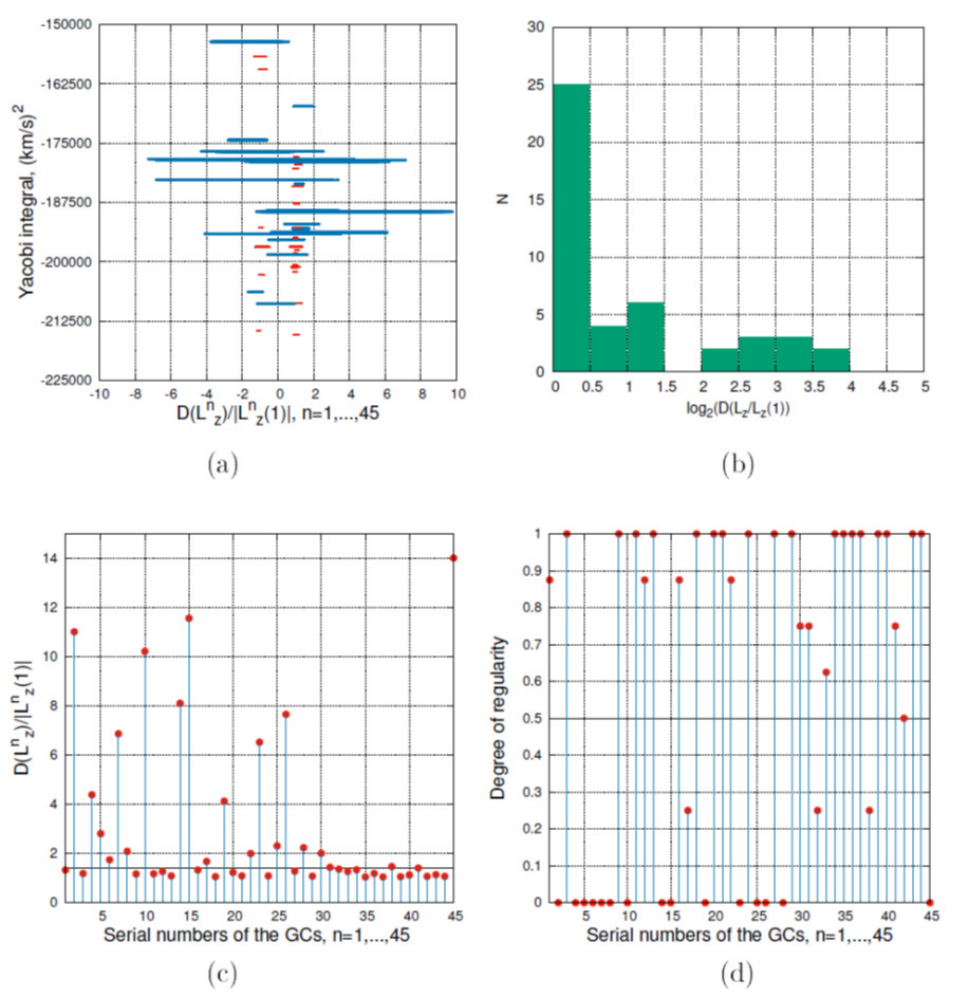}
         \medskip
\caption{\small Graphic illustration of the new method for determining the orbital dynamics of a GC based on the magnitude of the variation $D(L_z/L_z(1))$: (a) diagram $D(L^n_z/|L_z^n(1)|$ -- Jacobi integral", (b) histogram of the distribution of variations $D(L^n_z/|L_z^n(1)|$, (c) values of variations in $D(L^n_z/|L_z^n(1)|$ for all 45 GCs, (d) degree of orbital regularity of all 45 GCs, determined by the "voting" method.}
    \label{fig:f1}
\end{center}}
\end{figure*}

\end{document}